\documentclass{article}
\usepackage{spconf,amsmath,graphicx}
\usepackage{amssymb}
\usepackage{amsmath}
\usepackage{color}
\usepackage{subcaption}
\usepackage[hyphens]{url}

\newcommand{\ignore}[1]{}

\title{An Experimental Analysis of the Power Consumption of\\
  Convolutional Neural Networks for Keyword Spotting}

\name{Raphael Tang \qquad Weijie Wang \qquad Zhucheng Tu \qquad Jimmy Lin}
\address{David R. Cheriton School of Computer Science\\
	University of Waterloo\\
	\texttt{\{r33tang,w268wang,michael.tu,jimmylin\}@uwaterloo.ca}}

\begin{document}

\maketitle

\begin{abstract}
Nearly all previous work on small-footprint keyword spotting with
neural networks quantify model footprint in terms of the number of
parameters and multiply operations for a feedforward inference pass. These values
are, however, proxy measures since empirical performance in actual
deployments is determined by many factors. In this paper, we study the
power consumption of a family of convolutional neural networks for
keyword spotting on a Raspberry Pi. We find that both proxies are good
predictors of energy usage, although the number of multiplies is more
predictive than the number of model parameters. We also confirm that models
with the highest accuracies are, unsurprisingly, the most power hungry.
\end{abstract}


\maketitle

\section{Introduction}

Conversational agents that offer speech-based interfaces are
increasingly part of our daily lives, both embodied in mobile phones
as well as standalone consumer devices for the home. Prominent
examples include Google's Assistant, Apple's Siri, Amazon's Alexa, and
Microsoft's Cortana. Due to model complexity and computational
requirements, full speech recognition is typically performed in the
cloud:\ recorded audio is transferred to a datacenter for
processing. For both practical and privacy concerns, devices usually
perform keyword spotting locally to detect a trigger phrase such as
``hey Siri'', which provides an explicit acknowledgment that subsequent audio
recordings of user utterances will be sent to backend servers and thus
may be logged and analyzed. Beyond detecting these triggers,
it makes sense to perform recognition of simple commands such as
``go'' and ``stop'' as well as common responses such as ``yes'' and
``no'' directly on-device. Together, these represent instances of the
keyword spotting task on continuous speech input. Due to power
constraints on mobile devices, it is desirable that such keyword
spotting models are ``compact'' and have a ``small footprint'' (which
we formally define below).

Over the past several years, neural networks have been successfully
applied to the keyword spotting task (see more details in
Section~\ref{section:related}). When discussing the ``footprint'' of a
model, the literature usually refers to two easily quantifiable
values:\ the number of model parameters and the number of
multiplies for a feedforward inference pass. Model ``compactness'' is thus measured in
terms of these two quantities, which are of course proxies at
best. Ultimately, what matters most is the energy consumption during inference.

To our knowledge, previous work in keyword spotting stops short of
actual energy measurements. Thus, the primary contribution of this
paper is the deployment of a number of convolutional neural networks for
keyword spotting on a Raspberry Pi, where we are able to measure the
energy usage of various models and relate these measurements back to
the proxies used in previous work. We find that the number of multiplies
does indeed predict energy usage and model latency, as does the
number of parameters (albeit the relationship is weaker).
Therefore, in the absence of actual power measurements, these proxies
can be helpful in guiding model development,
although we advise caution in interpreting both measures.
Finally, as expected, we confirm that the most accurate models are
also the most power hungry, suggesting unavoidable tradeoffs with
this family of CNN architectures.

\section{Related Work}
\label{section:related}

The application of neural networks to keyword spotting, of course, is
not new. Chen et al.~\cite{keyworddnn} introduced multi-layer
perceptrons as an alternative to HMM-based
approaches. Sainath and Parada~\cite{keywordcnn} built on that work
and achieved better results using convolutional neural networks
(CNNs). They specifically cited reduced model footprints (for
low-power applications) as a major motivation in moving to CNNs.

Despite more recent work in applying recurrent neural networks to the
keyword spotting task~\cite{keywordrnn,SunMing_etal_2017}, we focus on
the family of CNN models for several reasons. CNNs today remain the
standard baseline for small-footprint keyword spotting---they have a
straightforward architecture, are relatively easy to tune, and have
implementations in multiple deep learning frameworks.

In this paper, we do not propose any new models for keyword
spotting. Instead, we conducted a thorough experimental analysis of the
power consumption of CNNs proposed by Sainath and
Parada~\cite{keywordcnn}, using Google's recently-released Speech
Commands Dataset~\cite{dataset} as the benchmark.

Canziani et al. \cite{canziani2016analysis} previously studied the
relationship between power consumption, inference time, and accuracy
using deep neural networks for computer vision tasks running on an
NVIDIA Jetson TX1 board. This work studies many of these same
relationships for keyword spotting, but on a more wimpy device, the
Raspberry Pi.

\section{Experimental Design}

All experiments described in this paper were conducted with Honk, our
open-source PyTorch reimplementation of public TensorFlow keyword
spotting models, which are in turn based on the work of Sainath and
Parada~\cite{keywordcnn}. We have confirmed that our PyTorch
implementation achieves the same accuracy as the original TensorFlow
references~\cite{honk}. All our code is available on
GitHub\footnote{https://github.com/castorini/honk} for others to
build upon.

\subsection{Model Description}

For feature extraction, we first apply a band-pass filter of 20Hz/4kHz
to the input audio to reduce noise. Forty-dimensional Mel-Frequency
Cepstrum Coefficient (MFCC) frames are then constructed and stacked
using a 30ms window and a 10ms frame shift. All frames are stacked
across a 1s interval to form the two-dimensional input to our models.

\begin{figure}
	\centering
	\includegraphics[width=0.45\textwidth]{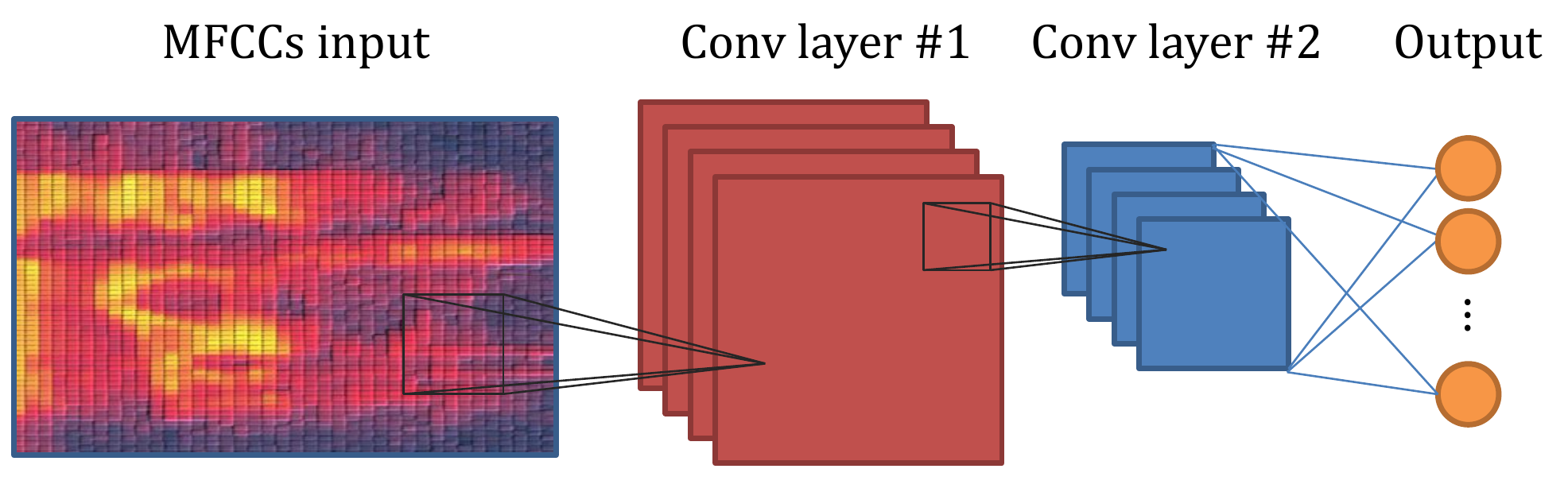}
	\caption{Convolutional neural network architecture for
		keyword spotting.}\label{fig:convnet}
        \vspace{0.2cm}
\end{figure}

\begin{table}
	\begin{center}
		\begin{tabular}{ r | c c c c c | c c}
			\hline
			type & $m$ & $r$ & $n$ & $p$ & $q$ & Par. & Mult.\\
			\hline
			conv & 20 & 8 & 64 & 1 & 3 & 10.2K & 27.7M \\
			conv & 10 & 4 & 64 & 1 & 1 & 164K & 95.7M \\
			lin & - & - & 32 & - & - & 1.20M & 1.20M \\
			dnn & - & - & 128 & - & - & 4.1K & 4.1K \\
			softmax & - & - & $n_{\text{labels}}$ & - & - & 1.54K & 1.54K \\
			\hline
			\hline
			Total & - & - & - & - & - & 1.37M & 125M\\
			\hline
		\end{tabular}
	\end{center}
        \vspace{-0.35cm}
	\caption{Structure of the \texttt{cnn-trad-fpool3} model.}
	\label{table:arch-trad-fpool3}
        \vspace{-0.2cm}
\end{table}

\begin{table}
	\begin{center}
		\begin{tabular}{ r | c c c c c | c c}
			\hline
			type & $m$ & $r$ & $n$ & $p$ & $q$ & Par. & Mult.\\
			\hline
			conv & 21 & 8 & 94 & 2 & 3 & 15.8K & 42.2M \\
			conv & 6 & 4 & 94 & 1 & 1 & 212K & 60.2M \\
			lin & - & - & 32 & - & - & 854K & 854K \\
			dnn & - & - & 128 & - & - & 4.1K & 4.1K \\
			softmax & - & - & $n_{\text{labels}}$ & - & - & 1.54K & 1.54K \\
			\hline
			\hline
			Total & - & - & - & - & - & 1.09M & 103M\\
			\hline
		\end{tabular}
	\end{center}
        \vspace{-0.35cm}
	\caption{Structure of the \texttt{cnn-tpool2} model.}
	\label{table:arch-tpool2}
        \vspace{0.2cm}
\end{table}

\begin{table}
	\setlength\tabcolsep{4.5pt}
	\begin{center}
		\begin{tabular}{ r | c c c c c c c | c c}
			\hline
			type & $m$ & $r$ & $n$ & $p$ & $q$ & $s$ & $v$ & Par. & Mult.\\
			\hline
			conv & 101 & 8 & 186 & 1 & 1 & 1 & 1 & 150K & 4.99M \\
			dnn & - & - & 128 & - & - & - & - & 786K & 786K \\
			dnn & - & - & 128 & - & - & - & - & 16.4K & 16.4K \\
			softmax & - & - & $n_{\text{labels}}$ & - & - & - & - & 1.54K &
			1.54K \\
			\hline
			\hline
			Total & - & - & - & - & - & - & - & 954K & 5.76M\\
			\hline
		\end{tabular}
	\end{center}
        \vspace{-0.35cm}
	\caption{Structure of the \texttt{cnn-one-stride1} model.}
	\label{table:one-fstride1}
        \vspace{-0.2cm}
\end{table}

The basic model architecture for keyword spotting, shown in
Figure~\ref{fig:convnet},
comprises one or more convolutional layers followed
by fully-connected hidden layers, ending with a softmax output.
More specifically, an input of MFCCs $\mathbf{X} \in \mathbb{R}^{t\times f}$ is
convolved with weights from the first convolutional layer, $\mathbf{W} \in
\mathbb{R}^{m \times r \times n}$, where $t$ and $f$ are the lengths
in time and frequency, $m$ and $r$ are the width and height of the
convolution filter, and $n$ is the number of feature maps.
If desired, the convolution can stride by $s \times v$ and max-pool in $p
\times q$, parameters which also affect the compactness of the model.
Rectified linear units are used as the activation function for each
non-linear layer.

\begin{table*}[t]
	\begin{center}
		\begin{tabular}{ l r | r r | r r r}
			\hline
			Model & Test Accuracy & Par. & Mult. & Latency/q (ms) & Energy/q (mJ) & Peak Power (W)\\
			\hline
			\texttt{one-fstride4} & 70.28\% & 220K & 1.43M & 40 & 28 & 0.99\\
			\texttt{one-fstride8} & 67.90\% & 337K & 1.43M & 42 & 29 & 1.02\\
			\hline
			\texttt{one-stride1} & 77.06\% & 954K  & 5.76M  & 100 & 115 & 1.52\\
			\texttt{trad-pool2} & 87.51\% & 1.38M  & 98.8M  & 146 & 306 & 2.60\\
			\hline
			\texttt{tpool2} & 91.97\% & 1.09M & 103M & 204 & 384 & 2.21\\
			\texttt{tpool3} & 91.23\% & 823K & 73.7M & 159 & 279 & 2.16\\
			\hline
			\texttt{trad-fpool3} & 89.43\% & 1.37M & 125M & 227 & 431 & 2.20 \\
			\hline
			Feature extraction only & --- & --- & --- & 31 & 19 & 0.80 \\
			\hline
		\end{tabular}
	\end{center}
	\vspace{-0.45cm}
	\caption{Performance of CNN variants on the Raspberry Pi in
          terms of accuracy, footprint, latency, and power
          consumption. The compact model is \texttt{one-stride1} and
          the full model is \texttt{trad-pool2}. For reference, we
          also include a condition that only performs feature
          extraction. Energy calculations and peak power exclude idle
          power draw of 1.9W.}
	\label{table:power-results}
	\vspace{-0.2cm}
\end{table*}

From this basic design, Sainath and Parada~\cite{keywordcnn} proposed
a number of specific models. We evaluated the following:
\begin{itemize}
\setlength\itemsep{-0.25em}

\item {\tt trad-fpool3}:\ The base model, illustrated in
  Table~\ref{table:arch-trad-fpool3}, comprises two convolution layers
  followed by a linear layer, a hidden layer, and a final softmax
  layer. All other variants are derived from this model.

\item {\tt one-fstride\{4,8\}}:\ Limiting the number of multiplies and
  parameters, these are compact variants that stride in frequency and
  also use only one convolution layer. Sainath and
  Parada found that {\tt one-fstride4} performs
  better than {\tt one-fstride8}.

\item {\tt tpool\{2,3\}}:\ These are variants that pool in
  time. Sainath and Parada found that, depending on
  the task, {\tt tpool2} has performance equivalent to or better than
  {\tt trad-fpool3}. See Table~\ref{table:arch-tpool2} for the
  parameter breakdown of {\tt tpool2}.

\item {\tt trad-pool2}:\ TensorFlow's variant of the base
  model {\tt trad-fpool3}, with comparable accuracy, but using fewer
  multiplies.

\item {\tt one-stride1}:\ TensorFlow's compact variant of {\tt
  one-fstride4} (detailed in Table~\ref{table:one-fstride1}). It
  uses a standard striding of 1$\times$1 and thus has more parameters
  and multiplies, but achieves better accuracy.

\end{itemize}

\subsection{Model Export and Inference}

To run model inference on the Raspberry Pi, we exported Honk models
written and trained in PyTorch to Caffe2 using
ONNX,\footnote{http://onnx.ai/} the Open Neural Network Exchange
format used for interchanging models between different deep learning
frameworks. While PyTorch is good for research and rapidly iterating
on model architecture, it was not designed to serve models in
deployment settings, unlike Caffe2, which supports running deep learning models
on production servers as well as mobile devices and Raspbian.
This feature makes Caffe2 especially useful for evaluating keyword
spotting models in environments where they will actually be deployed.
The practice of building and training models in one framework and
running inference in another has been used in production at Facebook.
We built Caffe2 from source for Raspbian, with the
\texttt{-mfpu=neon} flag, to specify the use of NEON (ARM's Advanced SIMD) optimizations.
Our models use 32-bit floating point operations and Caffe2 implements convolutions using
the \texttt{im2col} approach.

Evaluation was performed on a Raspberry Pi 3 Model B (ARM Cortex-A53) running Raspbian
Stretch (4.9.41-v7+). 
On the Raspberry Pi, we run a Caffe2 service which imports
an ONNX model and performs inference (as described above). To capture power measurements, the
Raspberry Pi is plugged into a Watts Up Pro meter, which has a USB port from which
measurements can be programmatically read. Power measurements are
taken at a frequency of 1 Hz from an external laptop connected to the
meter. The length of each experimental trial is sufficiently long (on
the order of minutes)\ that this resolution yields reasonably
accurate measurements. During each experimental trial, a script on the
Raspberry Pi iterates through all keyword classes for a fixed model, calling an API
served by the laptop to start and stop measurements before and after
the Caffe2 service call. Each Caffe2 service call evaluates all test examples
for a given keyword class. There were a total of 2,567 test examples for
all keyword classes combined.

\section{Experimental Results}

\begin{figure*}[t]
	\centering
	\includegraphics[width=0.45\textwidth]{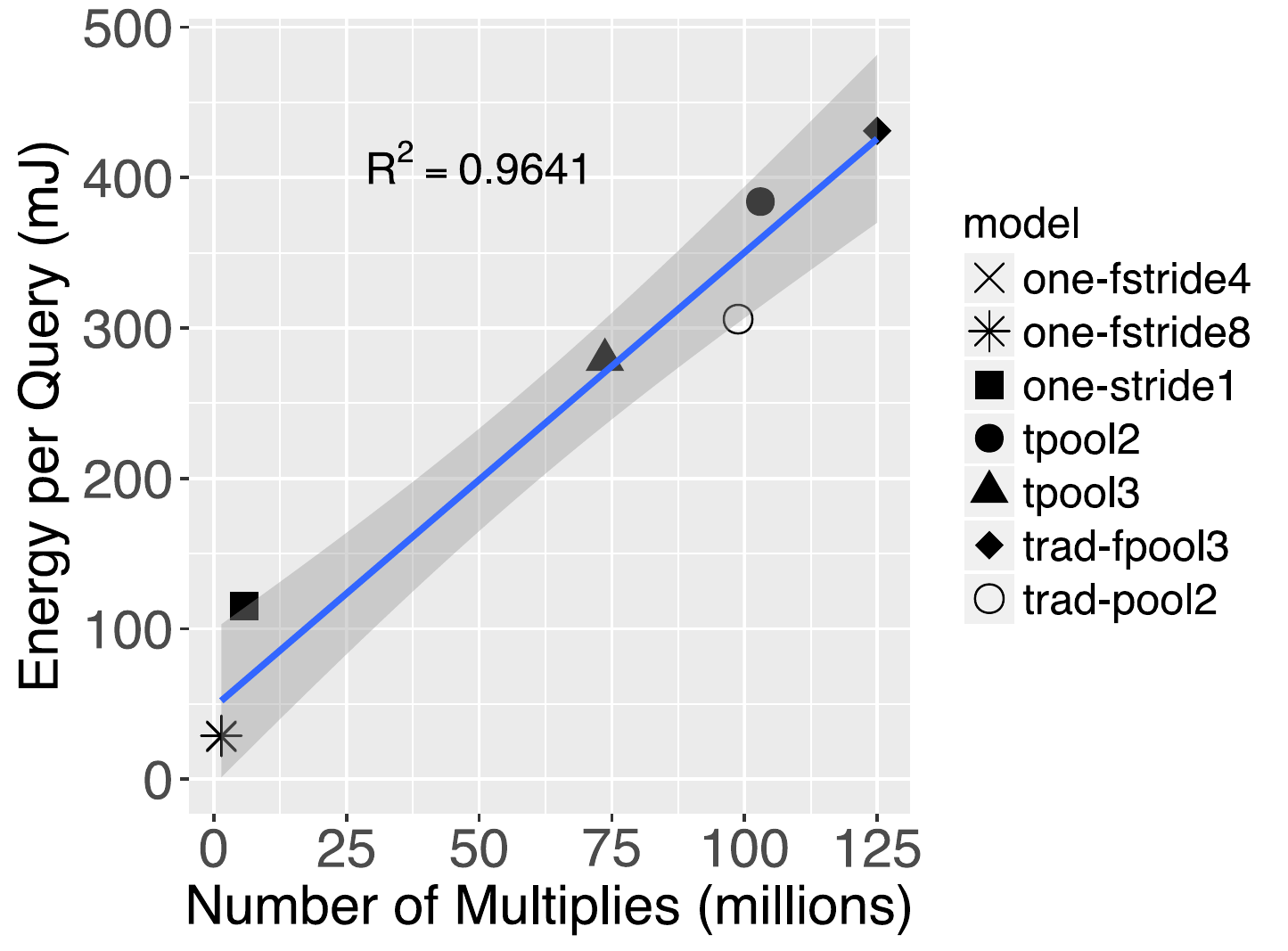}
	\includegraphics[width=0.45\textwidth]{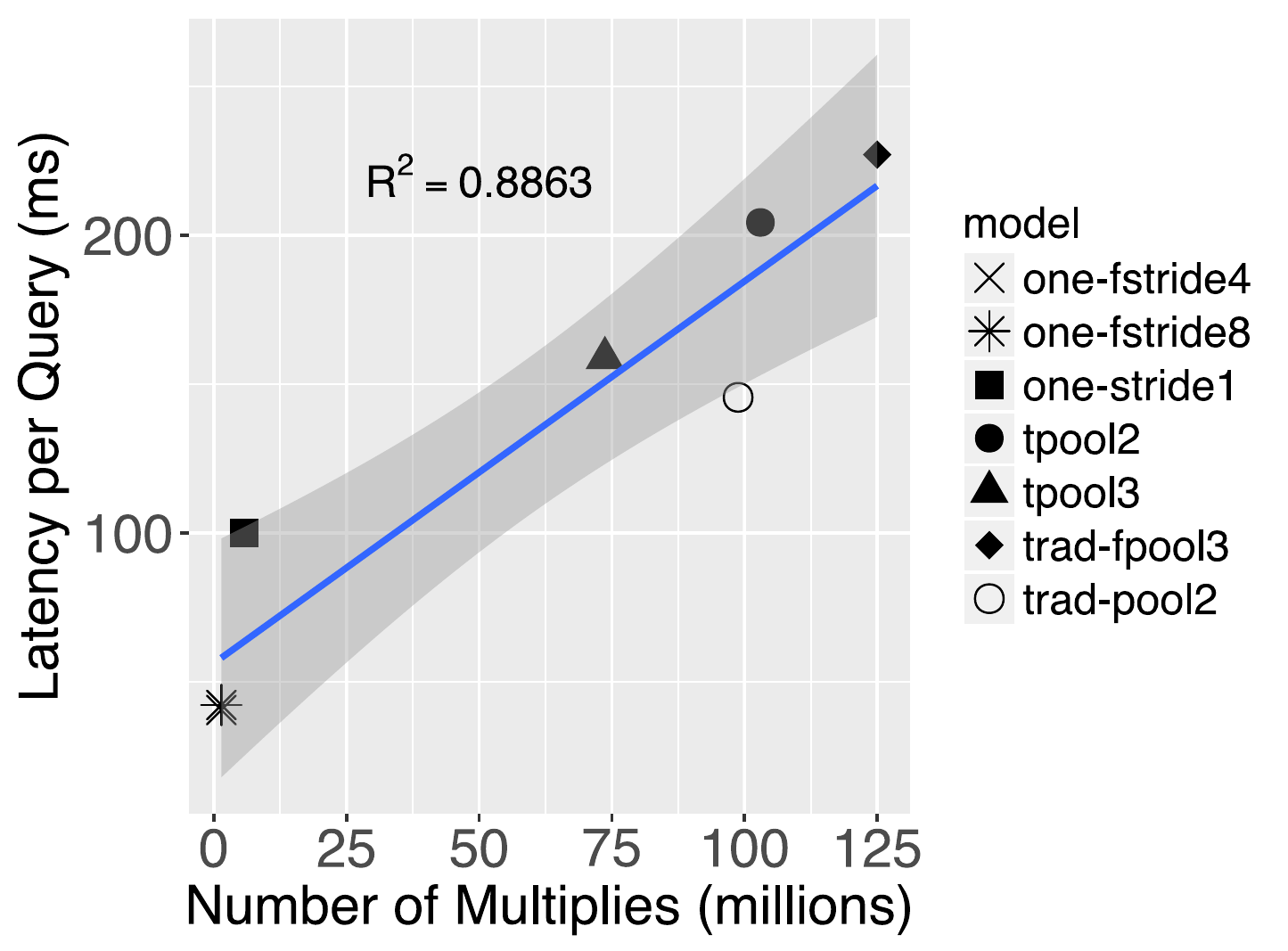}
	\caption{Energy (left) and latency (right) per query vs.\
          number of multiplies, with the 95\% confidence
          interval.}\label{fig:energy-multiplies}
\end{figure*}

\begin{figure}[t]
	\centering
	\includegraphics[width=0.45\textwidth]{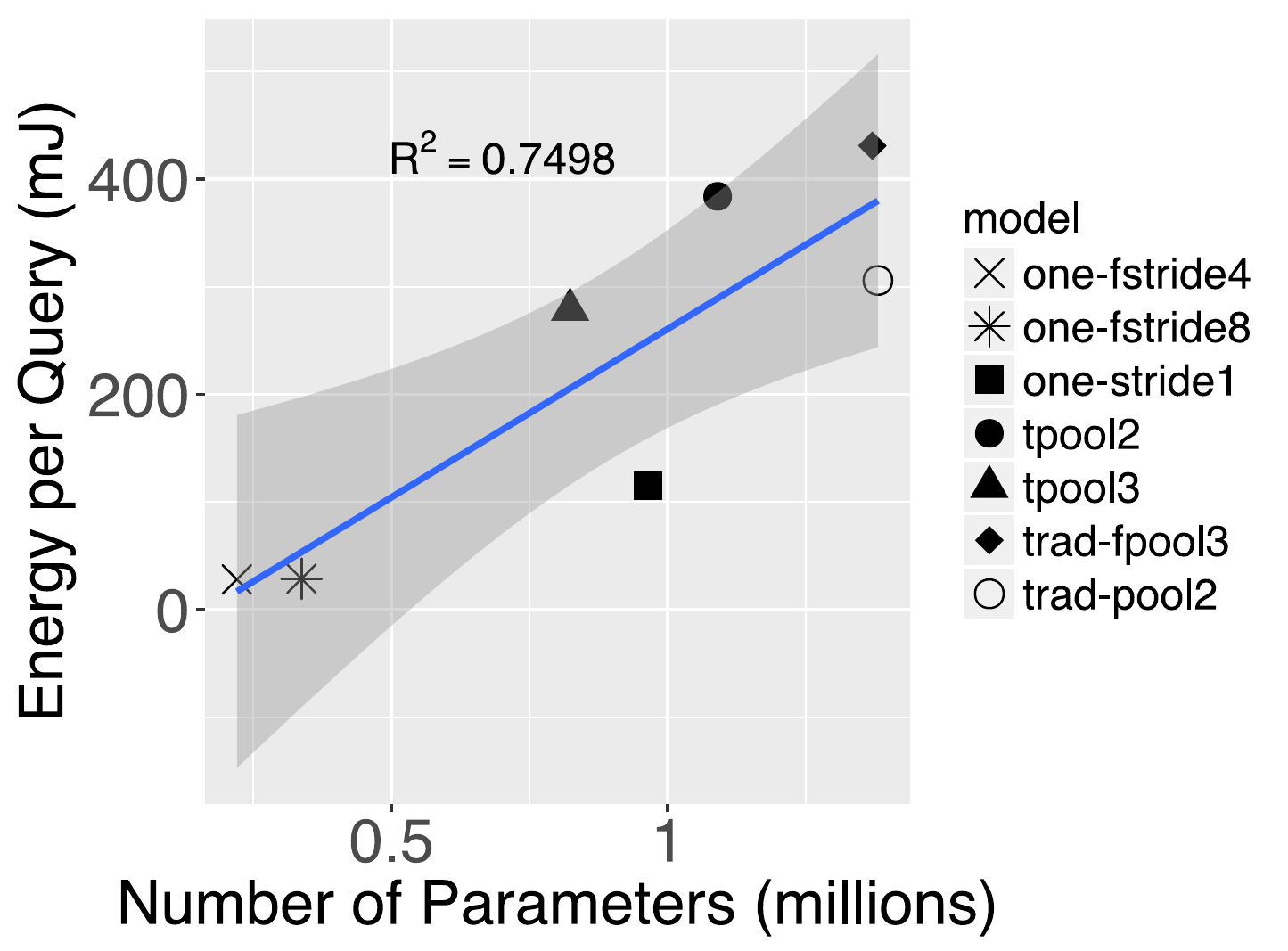}
	\caption{Energy per query vs.\ number of parameters (95\% CI).}\label{fig:energy-params}
\end{figure}

\begin{figure}[t]
	\centering
	\includegraphics[width=0.45\textwidth]{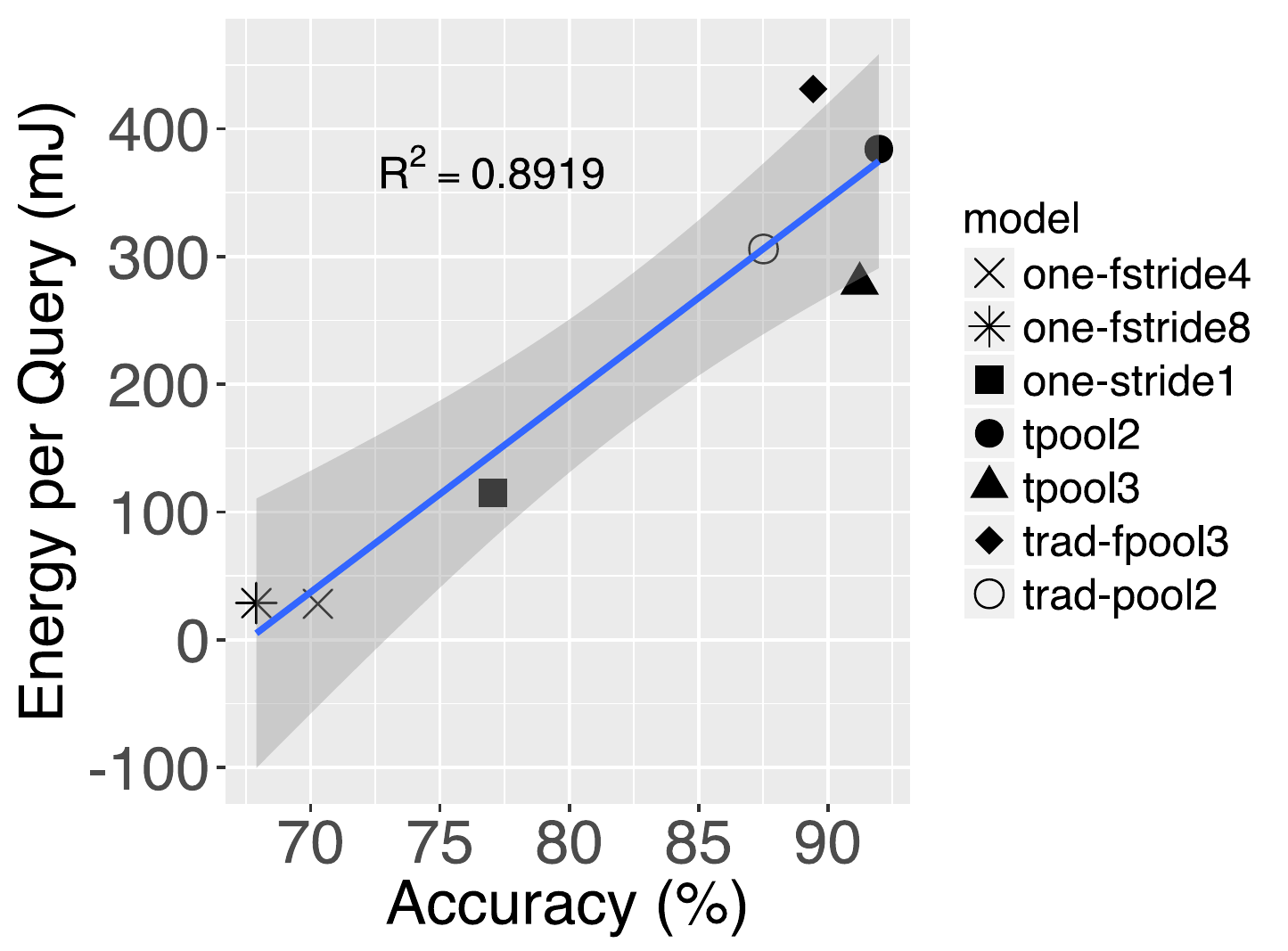}
	\caption{Energy per query vs.\ accuracy (95\% CI).}\label{fig:energy-accuracy}
\end{figure}

We evaluated the convolutional neural networks
described in the previous section using
Google's Speech Commands Dataset~\cite{dataset}, which was released in
August 2017 under a Creative Commons
license.\footnote{https://research.googleblog.com/2017/08/}
The dataset contains 65,000 one-second long utterances of 30 short
words by thousands of different people, as well as such background noise
samples as pink noise, white noise, and human-made sounds.

The Google blog post also references the
TensorFlow implementation of Sainath and Parada's models, which we
have ported to PyTorch and validated their correctness~\cite{honk}.
For consistency, we evaluated our PyTorch implementations,
but otherwise followed exactly the same experimental setup as
Google's reference. Specifically, our task is to classify a short
one-second utterance as ``yes'', ``no'', ``up'', ``down'', ``left'',
``right'', ``on'', ``off'', ``stop'', ``go'', silence, or unknown. As
the focus of this paper is on model performance in terms of energy
usage and not on accuracy per se, we refer interested readers to
details in Tang and Lin~\cite{honk}.
Our evaluation metric is accuracy, which is simply measured as
the fraction of classification decisions that are correct.

Our main results are shown in Table~\ref{table:power-results}. For
each model, we show its accuracy on the test set and its
model footprint in terms of the number of model parameters and the
number of multiplies required for a feedforward inference pass. The next
columns show the average query latency for each model on a test
instance, the energy per query, and the peak power draw during the experimental run.
The energy calculations as well as the peak power figures {\it exclude} energy consumed by the Raspberry Pi in
its idle state, which has a power draw of 1.9W. For reference, we also
report a condition that performs only feature extraction.

We found strong evidence of a positive linear relationship between the
number of multiply operations used in the models and the energy used per query
($R^2 = 0.9641$, $p = 0.0001$) in Figure~\ref{fig:energy-multiplies} (left) and also between the 
number
of multiplies and latency per query ($R^2 = 0.8863$, $p = 0.0015$) in
Figure~\ref{fig:energy-multiplies} (right). There is also strong evidence
of a positive relationship between the number of
parameters and the energy used per query ($R^2 = 0.7498$, $p = 0.0118$) in 
Figure~\ref{fig:energy-params}
and between the number of parameters and latency per query ($R^2 = 0.7237$, $p = 0.0152$), not 
shown.
However, the strength of correlations for the number of parameters is weaker.

These results suggest that the number of multiplies, and to a lesser
extent, the number of parameters, are useful proxy measures when
developing small-footprint keyword spotting models that optimize for
power consumption. Nevertheless, we suggest that
these metrics must still be interpreted with caution.
For example, we see that two models with similar numbers of multiplies
can still have very different energy profiles:\ \texttt{tpool2} and
\texttt{trad-pool2} have comparable numbers of multiplies but the
former is 40\% slower and consumes 25\% more energy per
query. However, the latter has a higher peak power draw.

Finally, we plot the relationship between energy usage and model
accuracy in Figure~\ref{fig:energy-accuracy}. The strong correlation
observed ($R^2 = 0.8919$, $p = 0.0014$) suggests that ``you get what you pay for'', in the sense that
at least for this family of models, a designer must trade off accuracy
for power consumption, and the relationship is surprisingly linear.

\section{Conclusions}

In this paper, we close a gap in the literature on small-footprint
keyword spotting. Previous work adopts the number of model parameters
and multiplies in an inference pass as optimization objectives, under
the assumption that smaller values translate into lower power
consumption. To our knowledge, this assumption has not actually been
verified until now. We do indeed find that both metrics are strong
predictors of energy usage, although the number of multiplies exhibits
a stronger correlation. While both are useful proxies during model
development, we noticed sufficient variations---for example, models
with similar multiplies but very different performance
profiles---that actual power measurements may still be required for
conclusive summative evaluations.

\end{document}